\documentclass[superscriptaddress,twocolumn,aps,prxlife,longbibliography]{revtex4-2}
\usepackage{graphicx}                   
\usepackage{hyperref}                   
\usepackage{amsmath,amssymb}             
\usepackage{multirow}	
\usepackage{amsmath}
\usepackage{amssymb}	
\usepackage{xcolor}	
\usepackage{amsmath,color}

\graphicspath{{figs/}}					

\DeclareMathAlphabet{\mathpzc}{OT1}{pzc}{m}{it}

\begin{document}
	
	\author{Hediye Yarahmadi}
	\affiliation{SISSA - Cognitive Neuroscience; Trieste, Italy}
	
	\author{Kwang Il Ryom}
    \affiliation{SISSA - Cognitive Neuroscience; Trieste, Italy}

    \author{Giuseppe Longobardi}
	\affiliation{University of York; UK}
	
	\author{Alessandro Treves}\email{(corresponding author: ale@sissa.it) }
	\affiliation{SISSA - Cognitive Neuroscience; Trieste, Italy}

	\title{Modeling Language Evolution Using a Spin Glass Approach}\date{\today}
	
\begin{abstract}
The evolution of natural languages poses a riddle to any theoretical perspective based on efficiency considerations. If languages are already optimally effective means of organization and communication of thought, why do they change? And if they are driven to become optimally effective in the future, why do they change so slowly, and why do they diversify, rather than converge towards an optimum? We look here at the hypothesis that disorder, rather than efficiency, may play a dominant role. 

Most traditional approaches to study diachronic language dynamics emphasize lexical data, but it would seem that a crucial contribution to the effectiveness of a thought-coding device is given by its core generative structure, i.e., its syntax. Based on the reduction of syntax to a set of binary syntactic parameters, we introduce here a model of natural language change in which diachronic dynamics stem from disordered interactions between/among parameters, even in the idealized limit of identical external inputs. We show in which region of ``phase space'' such dynamics show the glassy features that are observed in natural language across time. In particular, binary syntactic vectors remain trapped in glassy metastable (i.e., tendentially stable) states when the degree of asymmetry in the disordered interactions is below a critical value, consistent with studies of spin glasses with asymmetric interactions. We further show that an added Hopfield-type memory term, would indeed, if strong enough, stabilize syntactic configurations even above the critical value, but losing the multiplicity of stable states. Finally, using a notion of linguistic distance in syntactic state space we show that a phylogenetic signal may remain among related languages, despite their gradually divergent syntax, exactly as recently pointed out for real-world languages. These statistical results appear to generalize beyond the dataset of 94 syntactic parameters across 58 languages used in this study. 

\end{abstract}

\maketitle

\section{Introduction}

The study of languages is crucial for a profound comprehension of human cognition, culture, and history. Yet, the most basic observation, that natural languages are diverse, remains theoretically unexplained. As the myth of the Tower of Babel indicates, it would make sense for an effective means of coding thoughts to be common to all those who might need to share their mental states. Further, it would seem plausible that structurally different language forms would significantly differ in their efficiency, leading human groups to discard the less efficient ones. Yet, structurally very different natural languages survive and thrive to this day. Rather than converging under selective pressures towards a common standard, they even appear to diverge, albeit very slowly, as if drifting away from each other \cite{guardiano2005parametric,ceolin2021boundaries}. Drift dynamics might be expected to characterize superficial language features, such as the lexicon, but less so the internal syntactic structures. These structures are indeed more resistant to change \cite{keenan2014universal}, but they still change in the direction of increased diversity.

For two centuries, the study of language evolution has in fact largely relied on the form of lexical items \cite{swadesh1952lexico}. Recently, Bayesian phylogenetic methods have been applied to core vocabulary \cite{gray2003language,sagart2019dated,greenhill2020bayesian}, and geographic data has been integrated to model, e.g., the spread of Indo-European languages \cite{bouckaert2012mapping}. Historical constraints have been used to evaluate hypotheses on their place of origin \cite{chang2015ancestry}, and automatic word comparison methods have enhanced reproducibility and bridged traditional approaches with machine learning \cite{list2017potential}. 

Studying the lexicon, however, can contribute little to address the basic questions above, because words can change in a myriad ways, and it is not feasible to quantify and contrast any potential change in language efficiency with the number of possibilities effectively sampled, what in a physical system would be the energy-entropy balance. Furthermore, the study of lexical evolution has well known space-time limits defined by the concept of language family. Syntax, the structuring of words into sentences, is much more constrained: parametric approaches claim to reduce its diversity to a simple string of binary parameter values associated with each natural language \cite{chomsky1993theory}. The number of such parameters, initially thought to be few \cite{baker2008atoms}, has recently grown into the hundreds \cite{ceolin2021boundaries}, but they still span a much more limited set than the one spanned by lexical items \cite{moro2016impossible}.

In recent decades, in fact, phylogenetic linguistics has advanced through the application of quantitative methods to generative syntax and the comparison of parameter values 
\cite{roberts2007diachronic,biberauer2005changing,
longobardi2008syntactic,longobardi2009evidence, longobardi2013toward, galves2012parameter, ceolin2020formal, ceolin2021boundaries}. It has been shown that, contrary to long-held assumptions, syntactic features —- when analyzed within the generative biolinguistic framework —- can offer valuable insights into the long-term evolution of languages \cite{ceolin2020formal}. This approach reveals that syntax can carry a historical signal comparable to that of traditional etymological methods, but with sufficient verticality and greater chronological depth \cite{ceolin2020formal}. Recent research \cite{ceolin2021boundaries} has proposed that the integration of the Parametric Comparison Method (PCM) \cite{longobardi2009evidence, longobardi2013toward, ceolin2020formal} with theories of possible languages \cite{moro2016impossible} can elucidate phylogenetic relationships between language families. Following this line one may thus be able to extend the temporal scope of traditional comparative methods, offering a promising solution to long standing challenges in taxonomic linguistics \cite{ceolin2021boundaries}.

Chomsky’s original Principles $\&$ Parameters model describes grammatical diversity by attributing cross-linguistic variation to binary syntactic parameters.  It posits a universal and innate syntactic structure, with differences among languages arising from distinct parameter settings \cite{chomsky1993lectures,chomsky1993theory}, which are rapidly acquired during language learning by each individual. Language acquisition, in this view, thus involves selecting from a finite set of options \cite{fodor1998unambiguous}. More recent approaches within the minimalist program question the innateness of parameters, suggesting instead that they emerge dynamically through cognitive processes \cite{karimi2017introduction}. Still, aside from the innateness debate, linguistic variation is seen as governed by structured binary choices, offering a systematic account of phenomena such as variability in subject or object omission or word order. 

Syntactic parameter values cannot all be set, each independently of the others; parameters are interconnected through a hierarchical network of asymmetric implicational dependencies, exhibiting a near-logical structure shaped by functional constraints \cite{baker2008atoms,guardiano2016parameter}. For example, parameter B
may be defined only if parameter A takes value +1.
Seen the other way around, through the implications the value of one parameter may condition others. These can be regarded as strong, unidirectional interactions among parameters, that proceed always downstream from the top parameters, those that can be set independently of any other, toward the bottom of the hierarchy, occupied by parameters that can vary only subject to many conditions (see Fig.\ref{implications}, which refers to the parameters and their full implicational structure shown in Table S2 of \cite{ceolin2021boundaries}).
In a network model, such dependencies may be formalized through {\em polysynaptic} weights, as detailed below. 

Unlike biological evolution, where adaptive pressures can usually be identified that drive living systems toward greater efficiency (or {\em fitness}), syntactic change does not appear to follow such a straightforward route. While past syntactic states can often be reconstructed, the evolution of syntax -- lacking clear functional directions -- remains constrained but unpredictable. Its trajectories are marked by two seemingly contradictory features: persistence, where certain structures remain stable over time, and diversification, where languages diverge into distinct syntactic systems. 


The number and the complex interactions among syntactic parameters make statistical methods a reasonable option for their study. Although parameter values likely vary in subtle ways also across individuals nominally speaking the same natural language, the sets of parameter values of natural languages can be regarded as so-called {\em metastable} states. That is, they are attractive -- attractive to all the individuals sharing a particular language -- and not absolutely but largely stable, over centuries. The diversity expressed by the observed sets can be interpreted as stemming from the lack of any overarching order or functional principle, which might have imposed common values. This situation parallels glassy dynamics, in which numerous weak, disorderly interactions among components produce persistent, quasi-stable states that do not appear to conform to any ordering or symmetry principle. Therefore, disorder may play a pivotal role in syntactic evolution by enabling both diversification and persistence within language systems. 

It should be emphasized that these properties appear rather trivial if considered individually: diversification -- without attractor dynamics -- can be explained away by different external inputs, while persistence -- if diachronic changes were negligible -- is obviously a generic property of static systems. If syntax were, however, so directly sensitive to external inputs as required to produce diversification on its own, it would remain to be seen where it could find the stability necessary for speakers to share a grammar. In the most plausible case the relation between some hard-to-define primitive source in the external input and actual syntactic change is mediated by a number of internal factors: hence the necessity for first studying the overall functioning of such factors. This is an issue that has been avoided by theories of languages as synchronic codes or sets of rules, but has to be confronted by perspectives that view them as dynamical systems. To analyze, therefore, how diversification and persistence can coexist and interact, it seems necessary to go beyond linear reasoning and adopt a quantitative, multifactorial dynamical model. 

In the following, in Section II, we introduce a novel model of syntactic change inspired by glassy systems. This model combines strong, explicitly modeled asymmetric interactions that tendentially produce implicational relations, with a multitude of weak, statistically modeled pairwise interactions, expressing disorder in the exact way parameters influence each other. 
Already in its {\em vanilla} version, with purely binary parameters, the model can be used to probe the overall impact of these effectively random couplings on system dynamics. Section III presents a sharp transition obtained by slightly varying an angle defining the average asymmetry in the disordered interactions, and locates the transition point. In Section IV, we incorporate a {\em Hebbian} term as an additional symmetric interaction weight between parameters, and examine its effect on evolutionary dynamics. In Section V, we focus on language distances, analyzing the `historical signal', which reflects linguistic relatedness, Finally, in Section VI, we show that the above findings extend to a refined version of the model, in which implications determine whether a parameter can be defined or not, given the values of those above it in the hierarchy. Conclusions and future perspectives are summarized in Section VII.


\begin{figure*}
	\centering
	\includegraphics[width=17.8cm]{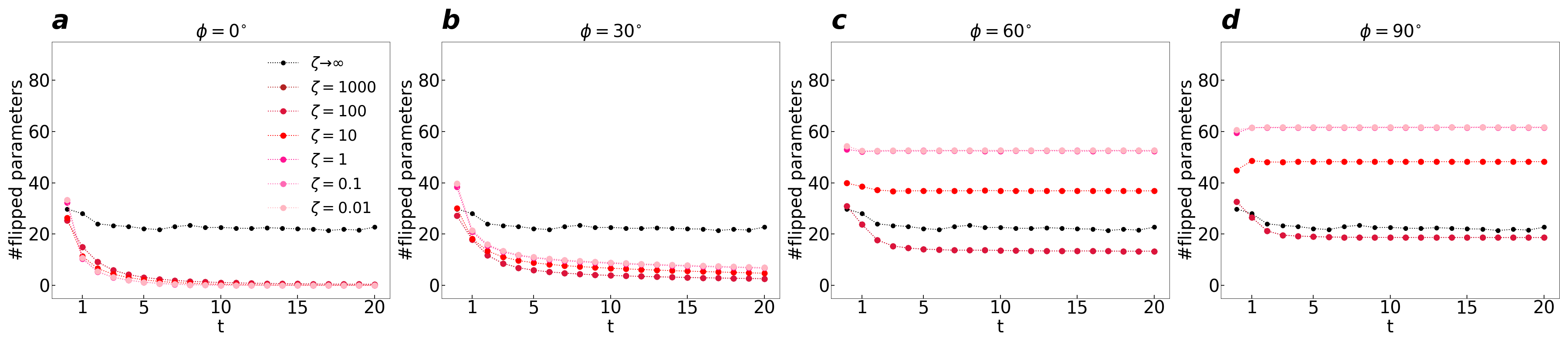}
	\caption{
		\textbf{Symmetric hidden couplings can block syntax evolution.} Simulations illustrating parameter fluidity as a function of implication strength \( \zeta \) and of the degree of symmetry  \( \phi \) in the hidden interactions. The black datapoints are the number of flipping parameters as a function of time for \( \zeta \to \infty \), i.e., with implications alone. Syntactic structures then continue to evolve, with 20–30 parameters flipping at each time step, indicating fluid dynamics. When adding hidden interactions, the strength \( \zeta = 0.01, 0.1, 1, 10, 100, 1000 \) of the implications is denoted by progressively darker colors.
        With hidden symmetric connections, (\( \phi = 0^{\circ} \), panel \textbf{a}), after the first few time steps, parameters stop changing, whatever the value of \( \zeta \). Eventually, after more time steps, the same {\em freezing} occurs in panel (\textbf{b}), with \( \phi = 30^{\circ} \); with more asymmetry, instead, i.e., \( \phi = 60^{\circ}\), (\textbf{c}) and \( \phi = 90^{\circ}\), (\textbf{d}), fluidity is maintained throughout, with the parameters continuing to flip over time, actually the more the weaker are the implications. What stabilizes very quickly is the average number of flipping parameter per time step. Each of 58 natural languages is used as a starting point for 10 simulations, for the same set of weak interaction weights; and then averages are taken over 100 distinct sets of weights. Therefore, the curves shown are averages over a total of 58,000 data points.} 
	\label{fig1}
\end{figure*}

\section{Modeling Syntactic Evolution}	
This study is based on simulating a simple model of the evolution of syntax over time. The model assumes that parameters can change value (or {\em flip}) under the influence of other parameters, and with negligible external inputs. Language change therefore proceeds through the attempt, by the parameters, to coordinate their values, in what is effectively a {\em recurrent} network. The crucial assumption is that complete coordination may never be possible, given the lack of intentional design: a large array of disorderly constraints tends to leave those mutually incompatible, to some extent, and prone to change. The assumption of negligible external inputs does not imply that we believe them to be irrelevant to real syntax dynamics, of course a multifactorial process; it is simply a methodological choice adopted with models of other complex systems addressing the same critical issue \cite{amit1989modeling}: do they allow for a multiplicity of attractor states even when isolated from external inputs? 

Internal influences may be of two distinct types: the {\em implications} cited above and {\em hidden} couplings, which represent all presumable interactions among parameters that have not yet been analyzed explicitly, and which therefore we treat statistically. The assumption is that parameters, as they are somehow implemented within the same brain, cannot be completely isolated from each other and so must interact in a multitude of ways, which we do not know.
The hidden couplings are randomly assigned in this statistical sense, but we assume them to be the same across all languages, as they should reflect common neural mechanisms, and we keep them fixed as syntax evolves in our simulations. We further assume in this study that each hidden coupling $W_{ij}$ involves only a pair of parameters, the influencer, $j$, and the influenced, $i$; a separate coupling $W_{ik}$ expresses the influence of another parameter $k$ on $i$, and yet another one $W_{ji}$ of $i$ on $j$. Non-pairwise or more complex hidden interactions are left for future studies. Importantly, we introduce an angle $\phi $ that statistically quantifies the degree of symmetry of such hidden couplings, from $\phi =0^{\circ} $ (fully symmetric couplings, e.g., from parameter $j$ to parameter $i$ and viceversa the two coupling coefficients would be equal, $W_{ij}= W_{ji}$); to $\phi =90^{\circ}$ (anti-symmetric couplings, i.e., they would be opposite, $W_{ij}=- W_{ji}$). 

Implications, on the other hand, explicitly reproduce linguistically derived conditions on parameter values, are inherently asymmetric and may involve multiple influencing parameters in a complex combination, e.g. parameter $i$ can be defined and hence may take its non-default value only if either parameter $j$ or parameter $k$ is set to +1. 

The model is applied to the most extensive available corpus of parametric syntax: the PCM dataset put together by Longobardi and collaborators \cite{longobardi2009evidence,ceolin2021boundaries}. It includes \(L=58\) languages from various families, each characterized by \(N=94\) syntactic parameters relative to the domain of nominal syntax. Each parameter value, denoted as \(\{S_{i}^{\mu}\}\), with \(i = 1, \ldots, N\) and \(\mu = 1, \ldots, L\), is set, if the parameter can be defined in that language, to either \(\{S_{i}^{\mu}\}=-1\) (default) or to \(\{S_{i}^{\mu}\}=+1\) (the {\em marked} value). Note that our notation aims to bridge with the physics literature on spin systems, hence binary values are identified with {\em S}pins \cite{sherrington1975solvable}; and with that on associative memories, hence distinct language configurations are indexed by the symbol $\mu$ \cite{amit1989modeling}. We refrain from using the symbol $p$ for either the number of {\em p}arameters considered or, as in the associative memory literature, the number of distinct {\em p}atterns or configurations of values ($=$languages) they are supposed to be able to express, and use $N$ and $L$ instead, respectively. 

In the PCM database, many parameters can also take the value 0, or {\em undefined}, if they are made undefined or irrelevant by the values of other parameters hierarchically above them. Here, we approach the issue of the undefined state in two different ways, as detailed below and in Sect.VI. Each language is then assigned a list of syntactic parameter values. The implications already limit the number of possible combinations the latter can take. Our goal is to study how the additional hidden couplings further affect changes in time of the vector of values associated with a community of speakers.

\subsection{The purely binary model}

In a first, basic version of our model, parameters can take only the marked value $+1$ or not take it, then remaining in the default state $-1$. In a later version, considered in Sect.VI, the unmarked value $-1$ is distinct from the undefined condition $0$, and parameters exert influences through the hidden couplings only when they are defined, in states $+1$ or $-1$.

In this first version, implications and interactions mediated by hidden couplings add up in the same equations, and the overall strength of the implications (relative to the hidden couplings) is scaled by a coefficient $\zeta $, which implies that for $\zeta \to\infty $ implications become absolute rules, while for $\zeta \to 0 $ they reduce to weak biases.


In summary, in this first version of the model we assume universal interactions \(\{J_{ij\dots k}\}\) between clusters of syntactic parameters, which are comprised in the most basic model of 
{\em I}mplications \(\{I_{ij\dots k}\}\) and of hidden pairwise couplings, or {\em W}eights \(\{W_{ij}\}\):
\begin{equation}\label{eq1}
J_{ij \dots k} =   W_{ij}+\zeta I_{ij \dots k},
\end{equation}
where \( \zeta \) denotes the overall strength of the explicit, sometimes complex implications, relative to that of the hidden, though simply dyadic interactions. Again, we use the notation \(\{J_{ij\dots k}\}\) for this set of universal conditions on parameter setting, for consistency with the statistical physics literature \cite{sherrington1975solvable}.

The dyadic interaction weights \(W_{ij}\) are expressed as a combination of symmetric and antisymmetric components:
\begin{equation}\label{eq2}
W_{ij} = \cos(\phi) W_{ij}^{S} + \sin(\phi) W_{ij}^{A},
\end{equation}
where both \(W_{ij}^{S}=W_{ji}^{S}\) and \(W_{ij}^{A}=-W_{ji}^{A}\) are taken to be random numbers drawn from the same normal (Gaussian) distribution of zero mean and unit variance. The variable \(\phi\), ranging as noted above from \( \phi = 0^{\circ}\) (fully symmetric weights) to \( \phi = 90^{\circ}\) (fully antisymmetric) is expected to play a crucial role, as symmetry generally tends to promote stability. 


In analogy with statistical physics models of glassy Glauber dynamics \cite{fisher1963lattice} we update the values of the syntactic parameters in discrete time by choosing at each time step a random sequence (a permutation $i_p(i)$ of the $N$ parameters) and using in turn the {\em zero temperature} equation (cp. Eq.(\ref{eq3simpler}) below)
\begin{equation}\label{eq3}
S^{\mu}_{i_p}(t+1) = \text{sign} \left[ \sum_{j \dots k} J_{ij \dots k} S^{\mu}_j(t') \dots S^{\mu}_k(t') - \theta_i \right],
\end{equation}
where \(S^{\mu}_j(t')\) represents the state of the \(j\)-th syntactic parameter in language \(\mu\) at time \(t\), if $j$ has not yet been updated, $i_p(j)>i_p(i)$, and at time $t+1$ if already updated; and \(\theta_i\) is a threshold controlling, in general, the likelihood that parameter 
\(i\) would remain in the unmarked state (\(S_i = -1\)). How should the thresholds \(\theta_i\) and the implication weights $I_{ij \dots k} $ be determined? In this first model, they are set so that, in the absence of the hidden couplings $ W_{ij}$, the sum $ \sum_{j \dots k} I_{ij \dots k} S^{\mu}_j(t') \dots S^{\mu}_k(t') - \theta_i $ is exactly zero when the other parameter values are such that parameter $i$ can choose state ($\pm 1$ with equal probability), and is negative in all other conditions (then forcing the value -1). As can be seen, this stipulation uniquely determines both \(\theta_i\) and $I_{ij \dots k} $.

In the more refined model of Sect.VI, we have also run simulations with non-zero temperature, in which essentially Eqs.(\ref{eq3}) are satisfied only with a certain probability, which decreases from 1 to 1/2 as the temperature increases. In that second version of the model, the implications are implemented as hard constraints, and the thresholds serve a different function, to set the propensity for each parameter to take the marked +1 value. These points will be taken up later.

\subsection{Symmetric hidden couplings stabilize syntax}

We begin by simulating the future evolution of each of the 58 natural languages in the database, 10 times, with each of 100 distinct hidden coupling sets, for a total of 1000 runs per language. We collect the number of parameters flipping at each time step and for a first statistical assessment we do not look at each language individually, but rather at averages over a total of 58,000 data points. The results are plotted in Fig. \ref{fig1}.

The black curve in Fig.~\ref{fig1} shows the number of parameters flipping over time, for \( \zeta \to \infty \), i.e. subject solely to implicational constraints, those typically emerging in linguistic analyses. Syntactic structures then continue to evolve rapidly, with 20--30 parameters flipping at each time step, which is obvious considering that some of the parameters in the PCM database are not constrained by implications (13 out of 94) and therefore flip with probability 1/2, enabling other parameters to also flip in their wake (see Fig.~\ref{figS1} in the Appendix). In this model, therefore, were it not for the hidden couplings, natural languages would be fluid, and change over a time scale set, essentially, by the time step. 

Introducing hidden symmetric connections (\( \phi = 0^\circ \)) constrains these dynamics: a relatively large number of parameters (e.g., 10--30) still flip during the first or first few time steps, after which the system rapidly gets immobilized, or {\em frozen } in a configuration from which it cannot escape (Fig. \ref{fig1}\textbf{a}), regardless of the exact strength of the implications, \( \zeta\). 

Obviously, freezing reflects the assumption, in this particular framework, that syntactic parameters are influenced solely by endogenous compatibility, in the absence of any external inputs. Consequently, the system rapidly attains a stable equilibrium state; even very weak unknown interactions (corresponding by complementarity to large \( \zeta\)) suffice to freeze its evolution. This scenario, albeit a limit case, is interesting insofar as it is not an automatic consequence of the assumptions in the model.

Introducing asymmetry in the hidden interactions, in fact, can restore fluidity to the syntax network. This is not yet evident for \( \phi = 30^{\circ} \) (Fig.\ref{fig1}\textbf{b}) but it is for higher asymmetry values, like \( \phi = 60^{\circ} \), Fig.\ref{fig1}\textbf{c}, all the way to full anti-symmetry, \(\phi =90^{\circ}\), Fig.\ref{fig1}\textbf{d}. 

Therefore, the way parameters interact with the hidden couplings has a dramatic effect on the system dynamics; and however weak they may be, the key factor is their degree of symmetry, not their strength.

Partitioning syntactic parameters into classes based on their dependency order -- the independent ones in class $D_0$, those dependent on a single other parameter in $D_1$, those on two, $D_2$, on three, $D_3$, and so forth, following the arrows shown in Fig.\ref{implications} -- allows a better understanding of how the degree of symmetry in the hidden couplings affects network dynamics. The implications create a hierarchy among parameters and, when they evolve subject to implications alone, those on top of the hierarchy are completely free, while those in stepwise lower classes are progressively more constrained, and only a smaller fraction flips at every time step. Random symmetric couplings, however weak, make all parameters freeze, irrespective of their position in the hierarchy: syntax evolution is blocked, obstructed by the multitude of mutual constraints (at zero temperature, i.e., in the absence of fast noise). When the hidden couplings are anti-symmetric, Fig.~\ref{figS1} shows that the original hierarchy is expressed again, in a quantitatively similar fashion, at least close to the deterministic limit (\(\zeta \) large). Again, the fluidity of the system relies heavily on that of the 13 central $D_0$ parameters, and lower classes are progressively more constrained, even though the hidden couplings are not informed by the class structure.


\begin{figure}
	\centering
	\includegraphics[width=7.5cm]{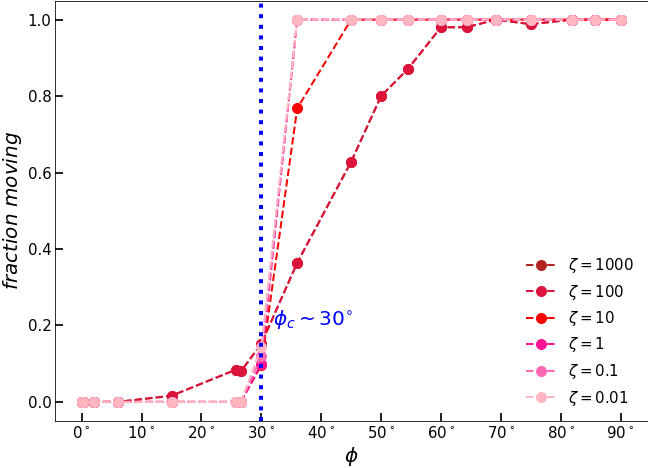}
	\caption{\textbf{Asymmetry makes the system fluid.} The fraction of parameter vectors still changing at $t=100$ as a function of the asymmetry angle \(\phi\). Data for 6 different \(\zeta\) values. The dashed line indicates the presumed critical value of \(\phi\), \(\phi_c = 30^{\circ}\), which is see to mark the crossing point of the curves for different values of \(\zeta\). A weaker implication strength results in a more abrupt transition. Note that datapoints for $\zeta = 1, 0.1$ and $0.01$. as well as for $\zeta = 100$ and $1000$, are superimposed onto each other, indicating that the critical $\zeta $ range is around $\zeta \simeq 10-50$.	} 
	\label{fig2}
\end{figure}

\begin{figure}
	\centering
	\includegraphics[width=9cm]{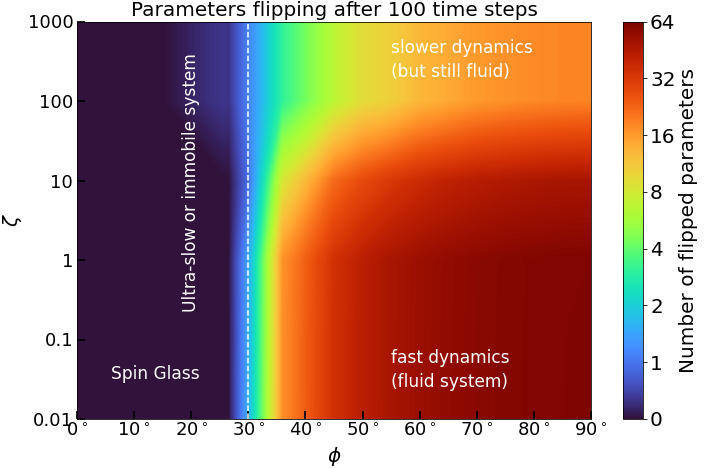}
	\caption{\textbf{Phase diagram for the basic model.}
The heatmap shows the mean number of parameters flipping at the $100^{th}$ time step, on a logarithmic color scale. The angle \(\phi\) (in degrees) is on the x-axis, and the strength of the implications \( \zeta \) is (logarithmically) on the y-axis. Towards the bottom of the diagram (small \(\zeta\)), implications become irrelevant, and the model behaves like a simple unstructured disordered system. For nearly symmetric interactions (up to \(\phi \approx 30^\circ\), marked by the white boundary), dynamics come to halt, as in a spin glass, while more asymmetric interactions lead to fluid behavior. At the top of the diagram, large \(\zeta\) and weak but sufficiently symmetric hidden interactions still stabilize parameters in metastable states of indefinite duration. By increasing asymmetry, the system transitions to a fluid state, though dynamics are slower compared to those occurring with weaker implications and similar degrees of asymmetry.
	} 
	\label{fig3}
\end{figure}

\section{A phase diagram for syntax change}

The fluidity of the system - whether the network of syntactic parameters evolves or not - appears to be relatively independent of the natural language considered, i.e., the starting binary vector of each simulation, but also, largely, of the strength $\zeta $ of the implications. The critical determinant of fluidity is the degree of symmetry $\phi $ of the hidden interactions. Its role can be highlighted by approaching the limit $\zeta \to 0$, in which implications become irrelevant. In such a limit, our model becomes an  asymmetric couplings variant of the Sherrington-Kirkpatrick model of a spin glass \cite{sherrington1975solvable}.

There have been relatively few studies of models of spin glass systems subject to asymmetric interactions \cite{nutzel1991length, nutzel1993subtle,crisanti1993transition}, and they differ in details like the statistics of the hidden couplings, the update rule or the parametrization of the degree of asymmetry. The couplings in these systems, \( J_{ij} \), are typically expressed as a combination of symmetric and antisymmetric components using a slightly different notation, $
J_{ij} = J_{ij}^{S} + k J_{ij}^{A}, $ ranging from strictly symmetric (\( k = 0 \)) to purely antisymmetric \( k \to \infty \). The symmetry factor $k$ is in turn related to the commonly used quantity \( \eta \), defined as
$
\eta := \langle J_{ij} J_{ji} \rangle/ \langle J_{ij}^2 \rangle,
$
which leads to the relations, useful to bridge across studies,
$
\tan \phi = k = \sqrt{(1 - \eta)/ (1 + \eta)}.
$
What is important is that these 3 formulations are alternative but equivalent: an average lack of either symmetry or anti-symmetry, that is, simple {\em a}symmetry, corresponds to $\phi =45^{\circ}$ or $k=1$ or $\eta =0$. 

It has been shown with a variety of approaches and in a number of slightly different variants that, as the degree of coupling asymmetry increases, the system crosses over from a glassy to a fluid regime. The transition may be observed by looking at cycle length \cite{nutzel1991length, nutzel1993subtle}, correlation entropy \cite{crisanti1993transition}, or relaxation time \cite{nutzel1991length, nutzel1993subtle,crisanti1993transition}. With nearly symmetric couplings, for instance, the cycle length remains independent of system size, while the relaxation time scales as a power of $N$. For highly asymmetric couplings, on the other hand, both cycle length and relaxation time grow exponentially with $N$. With parallel rather than sequential updating, Nutzel and Krey have identified a subtle transition point at $\eta = 0.5$ ($\phi=30^\circ$), where the attractor length changes significantly \cite{nutzel1991length, nutzel1993subtle}. 
For \( \eta > 0.5 \), the system gets trapped in fixed points or cycles of length 2. For \( \eta < 0.5 \) (more asymmetric couplings), the system enters chaotic dynamics, characterized by long cycles. Crisanti {\em et al}, on the other hand, have observed that the system becomes increasingly unpredictable once the asymmetry angle exceeds a critical threshold, and have proposed a scaling law for the mean Shannon entropy around $k_c\simeq 0.5$ (i.e., $\eta_c \simeq 0.6$, or in our terms $\phi_c\simeq 26^{\circ}-27^{\circ}$) \cite{crisanti1993transition}. They note that it could be a {\em bona fide} phase transition or a simple cross-over. Which of these two slightly divergent findings is in line with our own?

To better determine numerically the point at which the transition occurs, we count how often, on average across languages, more than 3 parameter still flip from the $99^{th}$ to the $100^{th}$ time step, at which point the average number of flips per time step is roughly constant. This analysis is based on 250 simulations per language (with 50 distinct weak coupling sets, each consisting of 5 runs). Then, we extract the fraction of moving parameter vectors (i.e., those that are flipping 3 or more parameters). Fig. \ref{fig2} shows that for decreasing values of \(\zeta\), the transition becomes progressively sharper, around the critical value \(\phi_c = 30^\circ \). Weaker implications bring the network of parameters closer to its statistical physics analog, and lead to a more abrupt transition, even keeping the same low $N=94$. The syntax network therefore shows, particularly clearly for small $\zeta$, a rather sharp transition, apparently for $\phi$ close to $30^{\circ}$, in agreement with the critical point proposed by Nutzel and Krey at $\eta_c = 0.5$ (or, equivalently, $k_c=1/\sqrt{3}$). 

Figure~\ref{fig3} shows a schematic phase diagram in the $\zeta$--$\phi$ plane, obtained by counting the average number of flipping parameters at the $100^{th}$ time step, i.e., in the asymptotic regime. The sharp change from a metastable to a fluid regime in the bottom part of the diagram is broadly consistent with the putative phase transition seen in asymmetric spin glass models the correspond to the limit $\zeta \to 0$ of our syntax network. As $\zeta $ increases, the crossover becomes smoother, but still remarkably well defined (considering also the small size of our system, $N=94$) and centered around the same value $\phi =30^\circ$. Note that along the vertical axis, the implication strength spans several orders of magnitude, on a logarithmic scale, and that a major decrease in fluidity is apparent, in the antisymmetric limit, only when $\zeta $ gets larger than $ \approx 20-50$. The factor $\zeta $ modulates the strength of individual implications, but since there are fewer implicational coefficients (a total of 753, 659 multi-unit coefficients $I_{ij\dots k}$ and 94 thresholds $\theta_i$) than pairwise couplings ($94\times 93=8742$, in our fully connected network) and they are also weaker, only above the range $\zeta \approx 20-50$ they can significantly constrain the fluid dynamics. Irrespective of $\zeta $, for sufficiently symmetric interactions, up to $\phi = 30^\circ$, the system exhibits glassy behavior with short-lived dynamics, and syntactic parameters rapidly settle into one of a disordered multitude of metastable states - a glassy phase. This parallels the slow dynamics and seemingly directionless diversification of grammars. The simulations then indicate that even weak yet sufficiently symmetric hidden interactions can stabilize multiple states. The behavior for large $\zeta$ is markedly different from the apparent asymptotic limit $\zeta \to \infty$, i.e., no hidden couplings, where syntactic parameters remain fluid, led by the 13 free ones among them. These findings suggest that incorporating disorder as an additional interaction term — alongside the intrinsic implications — plays a critical role in generating slow dynamics and long-term persistence in language evolution.




\begin{figure}
	\centering
	\includegraphics[width=9cm]{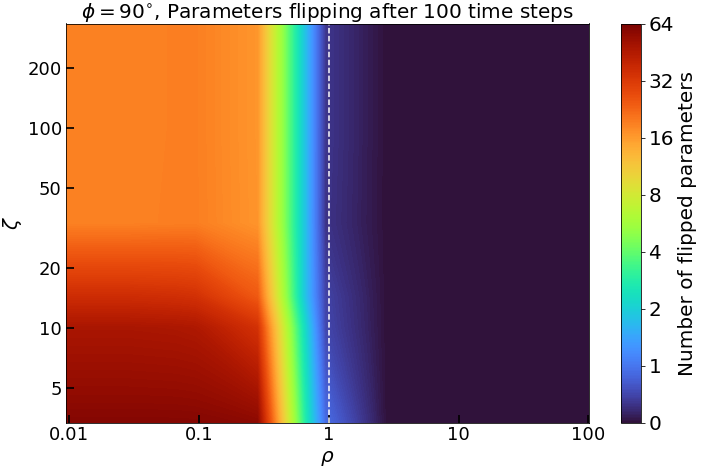}
	\caption{\textbf{A Hebbian term is effective only if strong.} Heatmap showing again the average number of parameters flipping at the $100^{th}$ time step, with fully antisymmetric hidden interactions (\(\phi = 90^\circ\)) when adding a stabilizing Hebbian term, as described in the text. The x-axis now denotes the factor \( \rho \), which regulates the strength of the Hopfield term, and the y-axis again \( \zeta \). Weak Hebbian interactions (up to $\rho \approx 1$), do not affect the fluid behavior prevailing with antisymmetric hidden interactions, as seen in Fig. 3. As the Hebbian interaction strength increases, the network becomes frozen. The cross-over appears rapid, but note the logarithmic $\rho $ scale.
	} 
	\label{fig4}
\end{figure}

\begin{figure}
	\centering
	\includegraphics[width=9cm]{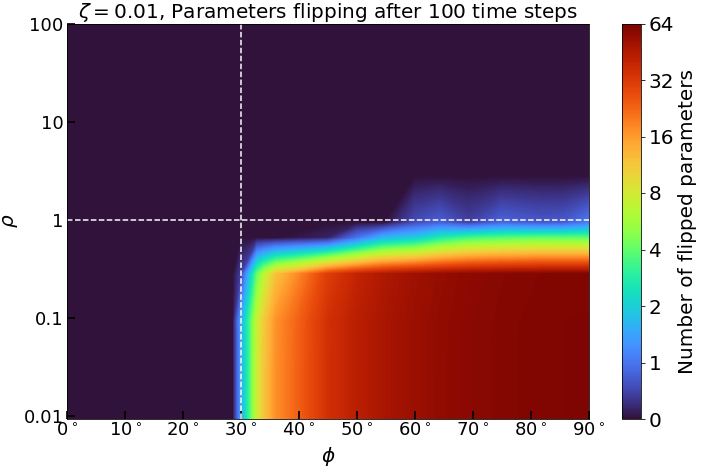}
	\caption{\textbf{An apparent OR rule for stabilizing syntax.} The heatmap with the number of flipping parameters at the $100^{th}$ time step, for small \( \zeta = 0.01\), in the \( \phi - \rho \) plane. Syntactic parameters stabilize (the blue region) if at least one of two conditions hold: sufficiently symmetric hidden interactions (up to $\phi \approx 30^\circ$) or a strong Hebbian term \( \rho \gtrsim 1 \). As discussed below, the two conditions are not however equivalent.
	} 
	\label{fig5}
\end{figure}

\section{Introducing a Hebbian term}

The model considered so far ignores the evidence provided by existing natural languages, in setting the hidden couplings - they are assigned purely at random. It may be hypothesized, however, that if specific realizations of syntax have arisen in human history, this may reflect interactions about particular syntactic parameters as they are implemented in the human brain. One simple-minded approach to delve into this complex and ill-defined issue is to introduce a set of {\em Hebbian autoassociative} synaptic weights, which in the syntax model would simply be additional pairwise interaction terms, strictly symmetric, that tend to stabilize existing realizations of language parameters. Inspired by the Hopfield model  of an autoassociative network \cite{hopfield1982neural}, which is known to preserve its glassy characteristics with moderate asymmetry \cite{treves1988metastable}, we may identify the 58 languages in the PCM database as the ‘memories' of the syntactic network. The new weights, denoted as \( W_{ij}^{\text{Hop}} \), to be included in the interaction term in Eq.~\ref{eq1},
$
J_{ij \dots k} = W_{ij}^{\text{RND}} + \zeta I_{ij \dots k} + \rho W_{ij}^{\text{Hop}},
$
are written as
$$
W_{ij}^{\text{Hop}} = 1/N \sum_\mu S^{\mu}_i S^{\mu}_j,
$$ 
where the factor \( \rho\) regulates their overall strength, \( S^\mu_i \) represents the binary state of syntactic parameter \( i \) for the \(\mu^{th}\) language, and  
\( \mu = 1, \ldots, L \) samples the available database. 



Figure~\ref{fig4} presents a schematic phase diagram in the $\zeta$-$\rho$ plane for a maximally fluid system -- one with fully antisymmetric hidden interactions ($\phi = 90^\circ$) -- in the  asymptotic regime (at the $100^{th}$ time step). Whatever the strength of the implications, there is clearly a cross-over from a fluid to a stable regime once the Hebbian term becomes effective, $\rho \gtrsim 1 $: the Hebbian term then stabilizes the system even when the other interactions are strictly antisymmetric. 

Figure~\ref{fig5} shows a similar phase diagram in the $\phi$-$\rho$ plane, when implications are negligible, $\zeta=0.01$ (note that $\rho$ has been moved to the y-axis). It appears to demonstrate that the stable regime, when syntactic parameters have completely stopped evolving by the $100^{th}$ time step, can be realized with either of two conditions, or both: when the Hebbian term is strong, $\rho > 1$, or hidden interactions are sufficiently symmetric, $\phi < 30^\circ$. The regime boundaries appear surprisingly close to a Boolean rule. 

\begin{figure}
	\centering	
    \includegraphics[width=9cm]{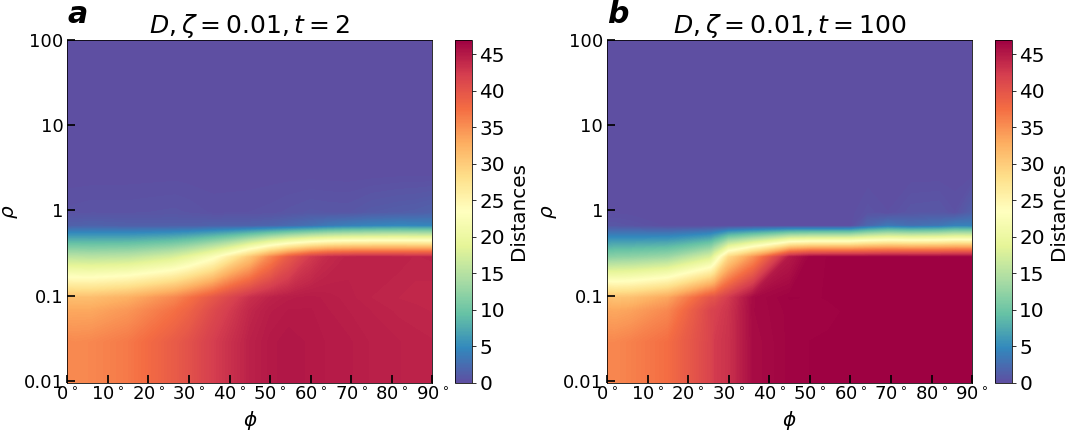}
	\caption{\textbf{Different languages rapidly converge, for $\rho > 1$.} Mean pairwise Hamming distances between realizations of distinct languages subject to identical hidden couplings, after only two time steps ((\( t = 2 \), panel (a)), and in the asymptotic regime (\( t = 100 \), panel (b)). These distances vanish when the Hebbian term is dominant (\( \rho > 1 \)), indicating rapid collapse, already after 2 time steps, onto a single syntactic structure. Note that in the fluid regime, bottom right corners, Hamming distances $\sim 47$ imply that typically half the parameters agree, between any two vectors, and half do not.}
	\label{fig6}
\end{figure}

In the high-$\rho $ regime, however, all syntactic configurations rapidly converge to a unique stable state, as shown in Fig.\ref{fig6}, which maps the average Hamming distance between vectors evolving from different starting configurations.


The reason for convergence to a single language, rather than to one of a disordered multitude of metastable states, is that unlike disorder, which halts parameter evolution through frustration, the Hopfield term, a form of order, suppresses linguistic diversity. The Hebbian term, therefore, does not seem able to account for the observed multiplicity of languages, when it dominates, and it does not lead to observable effects when it does not dominate. It could be, it should be noted, that a {\em weighted} contribution by the languages in the dataset, or the use of a {\em wider} sample of different languages, might each prevent the collapse and preserve multiplicity, i.e., language diversity. The available evidence is however insufficient to structure such differential weights.

In the remainder of this study, therefore, the Hebbian term will not be considered further, aside from a detail in Fig.~\ref{fig9}.

\begin{figure*}
	\centering
	\includegraphics[width=14.5cm]{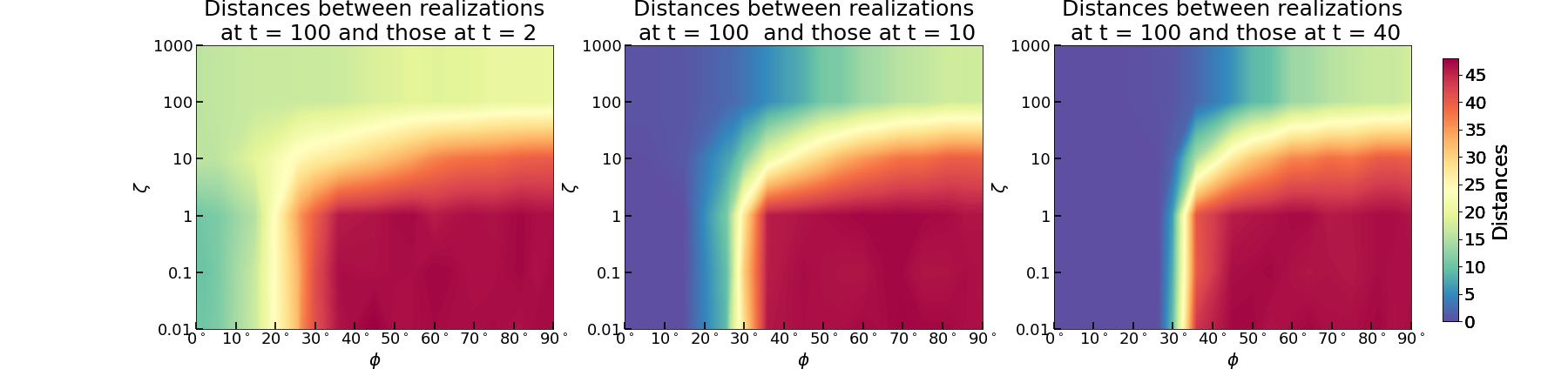}
	\caption{\textbf{Languages stabilize rapidly.}
    Hamming distance between realizations at the final time step (\( t = 100 \)) and those at earlier stages (\( t = 2 \), \( 10 \), and \( 40 \)) for Italian syntax, as an example. While such distance remains substantial in the region of high asymmetry to the right (larger when implications are weaker, bottom right), in prevailingly symmetric region in the left the syntactic vector has changed, at $t=100$, only if compared to very early stages of evolution, such as \( t = 2 \),
	(\textbf{a}). For \( \phi \) close to 0 it has already ceased to evolve by \( t = 10 \), (\textbf{b}), and definitely by \( t = 40 \), (\textbf{c}), all the way to nearly \( \phi \sim 30^\circ \).}
	\label{fig7}
\end{figure*}

\begin{figure*}
	\centering	\includegraphics[width=15.5cm]{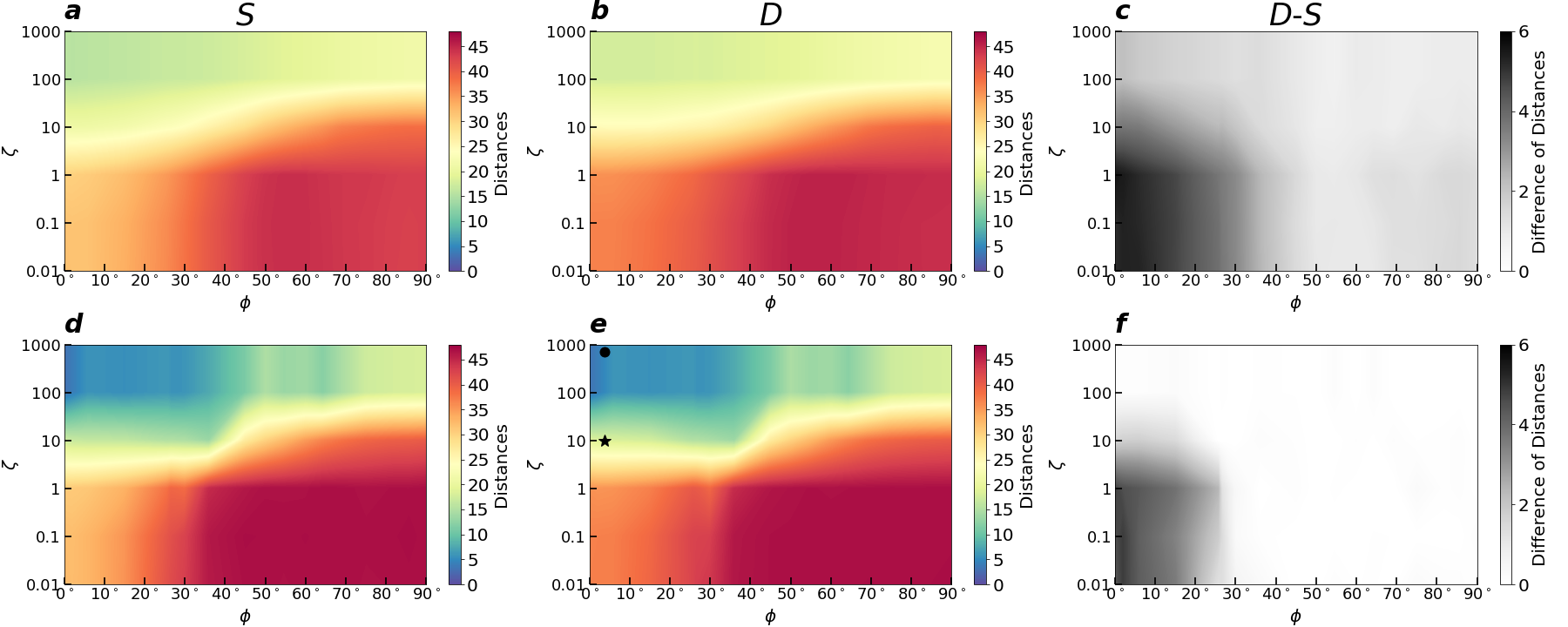}
	\caption{\textbf{Evolution of inter-language and intra-language distances}. The diagrams, in the $\zeta$--$\phi$ plane (at $\rho = 0$), show both at early ($t = 2$, top row) and asymptotic times ($t = 100$, bottom row) the average Hamming distance between simulations starting from the same natural language (distance in condition {\bf S}, left) and different natural languages (distance in condition {\bf D}, middle), as well as their difference, ({\bf D-S}, right), when using always identical hidden couplings. With time, both average distances diminish in areas characterized by strong implicational dynamics, especially when asymmetry is minimal. Still, they do not vanish: the average Hamming distance {\bf D} between descendants of different languages tends (asymptotically, or for $t=100$) to decrease from {\bf D} $\sim 35-37$ for weak implications ($\zeta <1$) thorough an intermediate range (e.g., {\bf D} $\sim 18$ for $\zeta =10$, the star symbol) down to values {\bf D} $\sim 3.1$ when implications dominate, for $\zeta > 100$ (round symbol). Their average difference {\bf D-S} illustrates the gradual fading of the phylogenetic/historical signal, i.e., the trace of shared ancestry among languages, everywhere except in the nearly symmetric and low (or at most intermediate) implicational strength regime, bottom left. Even there, the signal is however weak, a difference in at most $4-5$ parameters, compared to the dispersion observed starting from the same initial condition (left column).}
	\label{fig8}
\end{figure*}

\section{Preservation of a phylogenetic signal}

Tracing the persistence of syntactic structures in language evolution requires quantifying in some way how they diverge over time. To this end, we use as a first approximation the same simple notion of distance between languages used above to assess stability, i.e., the Hamming distance between different syntactic vectors - the number of parameters where the two vectors differ, ranging from 0 to 94 - where we do not rank syntactic parameters in terms of importance, or along their hierarchy. This metric ignores a wealth of linguistic knowledge, but can give a first impression, within the limits of our model, as to the long-term divergence or convergence among instantiations of syntax. We note that very similar observations can be based on the use of the Euclidean distance between vectors.


As an example of temporal evolution of a syntactic system within our model, we compare the presumed asymptotic state at \( t = 100 \) with selected intermediate stages (\( t = 2, 10, 40 \)) for the case of a real-world starting point — Italian — in Fig.~\ref{fig7}. When implications are irrelevant and asymmetry is high, the vector at $t=100$ remains very distant from its ancestors at essentially all preceding \( t \) stages -- note that a distance $\sim 47$ implies that half the 94 parameters take the same value in the two configurations, half a different value: effectively no syntax correlation. In other words, parameter values wander randomly in syntactic space, and reach no stable state nor do they approach an asymptotic state. Something similar occurs, albeit with reduced fluidity, when implications are stronger, but hidden couplings are still asymmetric. In contrast, the distance from the asymptotic configuration gradually vanishes if hidden couplings are prevailingly symmetric, and in fact when they are close to full symmetry syntactic parameters are already stable after a few time steps. These results are consistent across different languages.

If that is the time course of simulations starting from the same point, we need to compare them with those starting from other initial syntax vectors. A phylogenetic signal encodes the relatedness among descendants of the same syntactic configuration that persists, despite the initially rambling trajectories in syntactic space, produced by the hidden couplings. We obviously need to keep the assignment of hidden couplings the same, and compare distances between different runs with the Same starting point (case {\bf S}), and with Different starting points (case {\bf D}). Figure~\ref{fig8} presents this comparison, as a grand average, in the 
entire $\zeta$-$\phi$ plane (at $\rho = 0$), for configurations evaluated at two distinct time points: $t = 2$ (top row) and $t = 100$ (bottom row), corresponding to the early and asymptotic phases of the model, respectively. In each case, the main factor affecting these distances, which we label simply {\bf S} and {\bf D} like the cases they refer to, is the strength of the implications: when they are strong, they constrain the dynamics and both types of distance grow less - in fact, they decrease, reaching lower values at $t = 100$ than at $t = 2$. Thus, while for minimal asymmetry and in the low implication region, $\zeta < 1$, the average {\bf D} $\gtrsim 35$, in the strong implication regime, $\zeta > 100$, it goes down below one tenth of this value, {\bf D} $\sim 3.1$. When implications compete with the hidden couplings, $1<\zeta <100$, one observes intermediate values; but, whatever $\zeta $, there is always dispersal, i.e., a multiplicity of trajectories, which do not converge to a single language. 

Interestingly, however, when we take the difference {\bf S-D}, the only region where it remains above zero at $t = 100$ (even though with small values) is the region of nearly symmetric hidden couplings and weak or intermediate implications, up to about $\zeta \approx 50$. Strong implications, $\zeta > 100$, even with symmetric hidden couplings, retain a phylogenetic signal only for the first few time steps (top left region in the top right panel of Fig.~\ref{fig8}). This indicates that in this first, simply binary model the phylogenetic signal is washed away, after a sufficient number of time steps, if implications are strong.

 \begin{figure*}
	\centering
	\includegraphics[width=18cm]{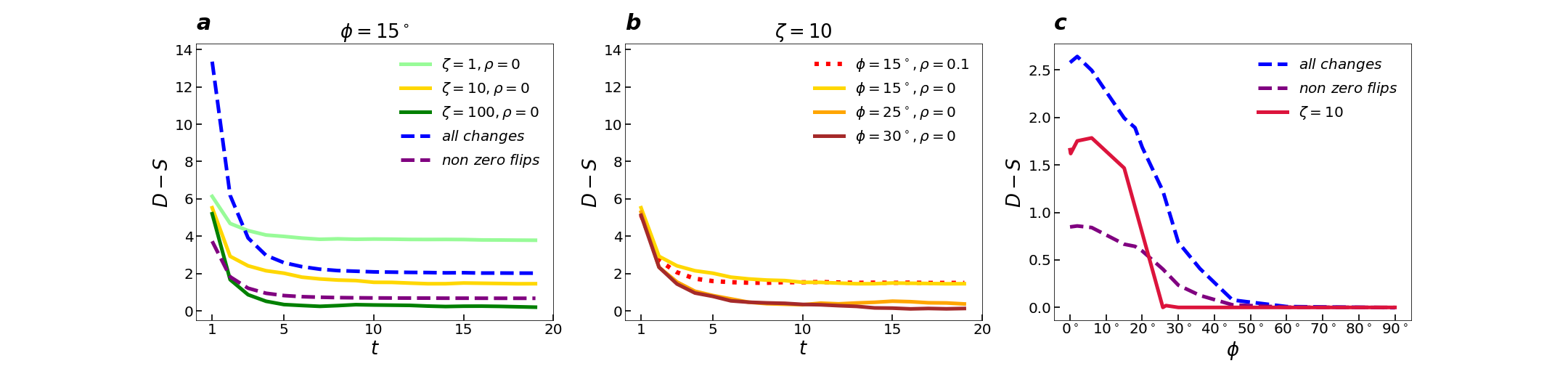}
\caption{\textbf{Temporal dynamics of the phylogenetic signal.}
	(\textbf{a}) The time course of the phylogenetic signal with $\phi$ fixed at $15^\circ$ and $\rho = 0$, showing the rapid decrease in signal followed by stabilization at a steady-state value. A robust asymptotic signal is still observed for $\zeta = 1$, i.e., when implications and hidden contributions would be balanced, except that implications are much sparser than the all-to-all hidden connectivity. As $\zeta$ increases, indicating a greater influence of implications over hidden terms, the steady-state signal decreases. In the refined model with the 0 state, however, the total signal (dashed blue curve) stabilizes at a level similar to that of the binary model with $\zeta = 10$.  
	(\textbf{b}) The impact of varying $\rho$ and $\phi$, while keeping $\zeta = 10$ constant. With limited asymmetry, $\phi = 15^\circ$, a small Hebbian term (the red dashed line for $\rho = 0.1$) leads the system to approximate the steady-state signal faster (compare to the yellow line for $\rho = 0$). Further asymmetry, i.e., increasing $\phi$ leads to the vanishing of the signal, concurrent with the cross over to the fluid regime.
    (\textbf{c}) The dependence on $\phi$ of the steady state signal (after 100 time steps) is similar, though not identical, for the model with the 0 state (total signal, dashed blue curve) as for the binary model (red curve).}        
	\label{fig9}
\end{figure*}


To further detail the time course of the average phylogenetic signal in the model, Figure~\ref{fig9} shows it as a function of $\zeta $ for $\phi =15^\circ$, in \textbf{a}, and as a function of $\phi $ for $\zeta =10$, in \textbf{b}. The signal rapidly decreases over a few iterations and then stabilizes at a steady-state value. This steady-state level depends on the variables $\phi$ and $\zeta$, and is stronger when they are both low. Notice in \textbf{b} that a weak additional Hebbian term, insufficient to lead to the collapse into a single stable state, does not alter the phylogenetic signal either, aside from initially accelerating the approach to the asymptotic state. Notice also that already at $\phi =30^\circ $ the signal has disappeared, as the syntactic network has crossed over into the fluid regime.

\section{The model with undefined parameters}

In the version of the model considered so far, implications are treated as contributing to the weighted input sum that determines the value of the affected parameter; they are strictly asymmetric and can involve multiple influences, but are otherwise on par with the hidden couplings. By decreasing their strength $\zeta$ they can be made into soft constraints or even slight biases, which enables the correspondence with the simple spin glass model in which they do not appear at all. It is possible, however, to implement them as precise, hard constraints that operate before ever taking the weighted sums. To that end, first we define a value 0 for the undefined state of parameters, distinct from their default value $-1$. 

 \begin{figure}
	\centering
	\includegraphics[width=9cm]{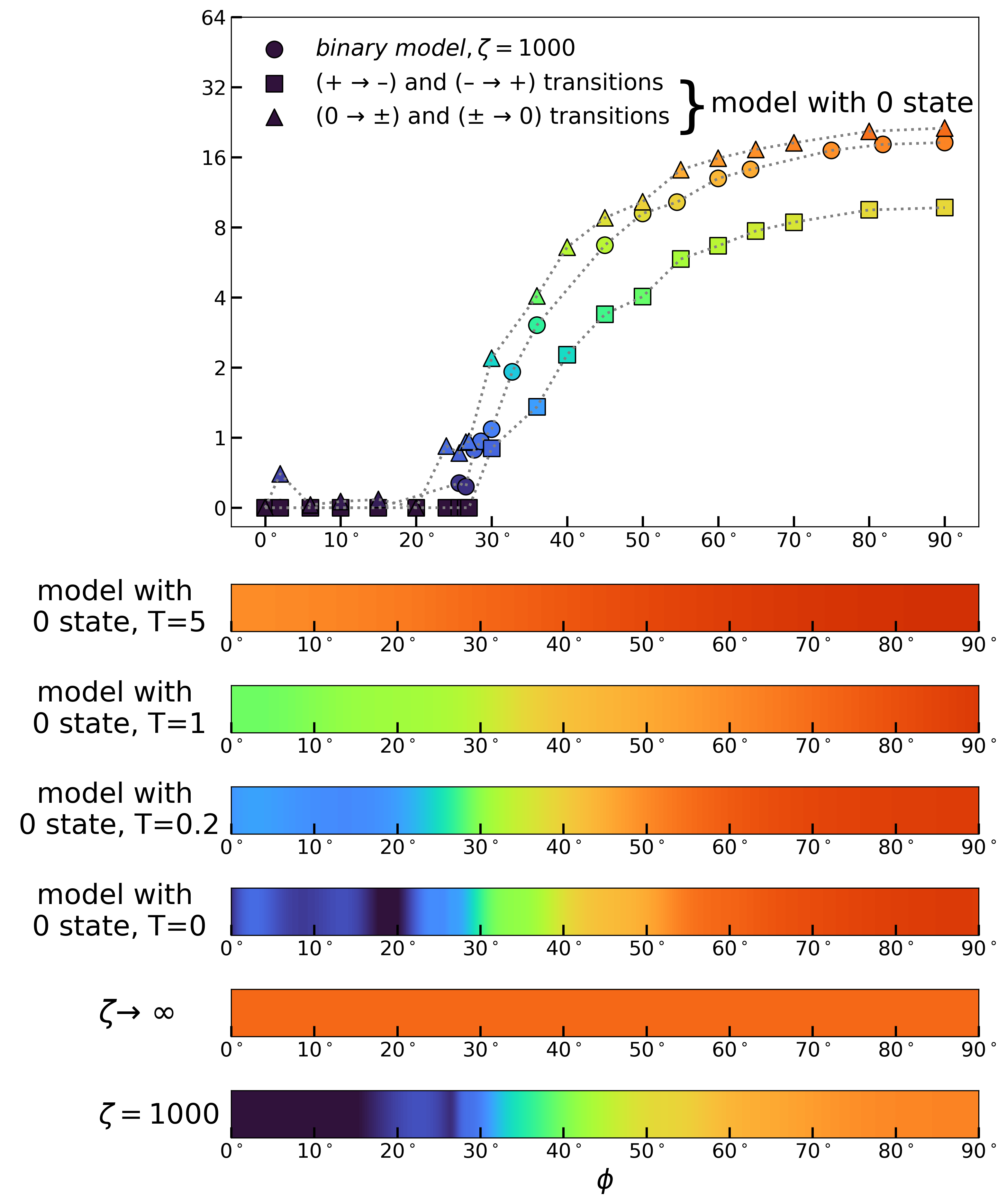}
\caption{\textbf{Fluidity in the model with the 0 state.}
	The number of parameters changing value at time step 100 is plotted (above) as a function of the asymmetry angle $\phi$, separately for the parameters that enter or leave the 0 state, and for those flipping between values $\pm 1$, and compared with those flipping in the binary model with $\zeta=1000$. The bars (below) represent the same data -- the sum of both types of change in the case of the revised model -- with the color coding of Fig.\ref{fig3}, and include results obtained with 3 non-zero $T$ values. The ‘hard constraints' model presents a similar $\phi-$dependence, for $T=0$, as the binary model with $\zeta=1000$, unlike the fluidity of the binary model with $\zeta\to\infty$. With higher $T$, the fluidity is gradually reinstated also for small $\phi$, but the model with $T \ll 1$ behaves similarly to the $T=0$ one: for $\phi < 30^\circ$ both types of change scale $\propto T$.
}
	\label{fig10}
\end{figure}

Partitioning parameters not in the classes $D_k$ mentioned above, but rather in levels (those at level $L_k$ are constrained by implications involving at least one parameter from level $L_{k-1}$, and in the PCM dataset $k=0,\dots,10$, see Fig.\ref{implications}), we then update their values going down the hierarchy, from $L_0$ to $L_{10}$, within each time step. For each level, each time step uses a different permutation of the parameters at that level, and considers them sequentially. First, it is checked whether the parameter is defined, according to the implications and to the values of the parameters above; if it is defined, it is updated using an equation with only the hidden couplings, setting it to the value +1 with probability
\begin{equation}\label{eq3simpler}
P(S^{\mu}_{i_p}(t+1)=1) =  \frac{1}{  1 + e^{-\left[ \sum_{j} W_{ij} S^{\mu}_j(t')- \theta_i\right]/T}}
\end{equation}
which when the ‘effective temperature' $T=0$ reduces to Eq.(\ref{eq3}) but without implications (or $\zeta=0$). When updating the value of one parameter at time step $t+1$, again we take the updated value at $t+1$ for all parameters above and those already updated at the same level, and the value at $t$ for those below and those at the same level not yet updated. Note that parameters in the undefined state (at $t'$) are set to the value 0 and do not contribute to the hidden interactions. The strength of the implications $\zeta$ becomes irrelevant: they are now effectively of infinite strength. The thresholds $\theta_i$ remain, but acquire a different function: they can now be used to bias the setting of the marked value +1, so that the probability with which it is taken by a given parameter approximates the frequency with which it is observed across natural languages. 

With these provisions, the dynamical behavior of the refined model remains similar to that of its simpler binary counterpart. This can be seen in Fig.\ref{fig10}: at $T=0$ the parameter network is trapped in glassy states for $\phi < 30^{\circ}$, like the binary model with finite $\zeta $. Increasing the ‘temperature' leads to fluid dynamics for all $\phi $ values, but gradually (the mean number of flips can be seen to be $\propto T$), unlike the sudden change of behaviour occurring when, in the binary model, the hidden couplings become strictly negligible with respect to the implications -- the $ \zeta\to\infty$ limit. The model with the 0 state is thus robust, i.e., its low-temperature dynamics is still glassy.  

\section{Conclusions and future perspectives}

This study presents a framework for modeling syntactic evolution through the lens of disordered systems, bridging formal linguistics and statistical physics. By treating syntactic parameters as binary variables subject to both logical and typological implications and to hidden interactions, the model aims to capture the dual behavior of syntactic systems: their remarkable persistence across time and their gradual yet unpredictable diversification.

Our results indeed indicate that in a portion of its phase space the network exhibits glassy dynamics, slow and gradually diverging, similar to that of spin glasses. Whether in the $N\to \infty $ limit the glassy regime would become a {\em bona fide} spin glass phase remains to be seen, but the glassy behavior is evident already in the network of limited size that models the $N=94$ syntactic parameters of the PCM database. Thus, the model suggests that syntactic evolution may not merely be the product of determinable, or even functionally driven processes, and might instead be shaped by subtle, usually unobservable interactions that may likely arise, e.g., from the way syntax is realized in the brain of the speakers. 
For $0<\phi < \phi_c = 30^\circ$ the isolated model would get stuck into a stable state, which would be only metastable, that is, slowly changing, in the realistic scenario in which external stimuli are taken into consideration; for $\phi > \phi_c = 30^\circ$, instead, syntax would continue to rapidly diversify, to the extent of losing any signal of persistence or indeed of realizing any shared syntactic rules even within the same community of speakers.
The critical asymmetry value $\phi_c = 30^\circ$, presumably marking a phase transition in the parallel statistical physics model, can serve as a valid reference point for the syntax model as well, as it is not much affected by other factors, in particular how the implications are modeled and, if they are taken as soft constraints, by their strength $\zeta $. Remarkably, the same value appears to hold if implications are taken to be hard constraints, in which case $\zeta $ becomes irrelevant. The suggestion from both versions of the syntax model, then, is that natural language parameters interact, perhaps due to their implementation in the brain, via couplings that are, if not strictly symmetric, at least prevailingly symmetric (quantitatively, $\phi < 30^{\circ}$).

Of course, in our idealized framework -- no external inputs whatsoever and zero or very low ‘temperature', i.e., minimal noise -- syntactic systems can become ‘frozen' in certain configurations, while in more realistic scenarios they would continue to evolve, though slowly, in the corresponding region of the phase diagram. It should be noted that a separate study has shown that a disordered network of binary variables interacting randomly with each other as well as with a set of multi-state variables (also randomly) can be even glassier, that is slower, than if it were isolated \cite{ryom2023speed}. The suggestion would then be that if syntax -- the binary parameters -- interacts randomly with e.g. the lexicon, or phonology, or semantics -- which can all be formulated in terms of multi-state variables -- it might evolve even slower than on its own. 

The introduction of a Hebbian memory term would stabilize syntactic configurations, even with asymmetric hidden couplings. However, an effective Hebbian term ($\rho > 1$) would lead all languages to collapse onto the same syntax, at least in our implementation of the Hebbian term. More sophisticated versions of the same idea might avoid the collapse. For example, it might be argued that treating closely related languages, like several of those from the Indo-European family that comprises nearly half of the current PCM database, as adding their separate Hebbian contributions might be excessive, theoretically dubious, and might be the key factor leading to the collapse. Still, it is not obvious at this stage how to construct a ‘balanced' Hebbian term, mimicking the effect of orthogonal memories in autoassociative memory networks \cite{hopfield1982neural}.

We have seen that in the first model, even in the absence of a Hebbian term, the phylogenetic signal rapidly vanishes if implications are sufficiently strong to really act as hard constraints, $\zeta\gtrsim 10-50$ (see Fig.~\ref{fig8}); this does not happen however in the more refined model with the zero or undefined state. Therefore, we can conclude  that indeed adding {\em weak} hidden interactions $W_{ij}$ to strong explicit implications suffices to lead to both divergence (the multiplicity of metastable states) and persistence (the phylogenetic signal). 

This study, in summary, aims to present the first steps towards mapping syntactic parameters onto a recurrent network model, and at a rather general level it can be said that model is compatible with empirical observations, {\em i.e.}, it fits the data. Can it also lead to predict more detailed data, yet to be collected? Clearly, before being able to make genuine predictions, the model needs to be developed in several directions. We mention here a few of them.

Two developments that are conceptually trivial, but might have a major practical effect, and which involve a lot of hard work, would be to extend the set of parameters and the set of languages now included in the PCM database. Current versions of the database include only Eurasian languages, and among them an over-representation of Indo-European ones. It is often assumed, but not really known, that the main statistical properties of their distribution of parameter values reflect more general trends among all world languages. Similarly, the database is restricted to parameters affecting the determiner phrase, perhaps half or less of those relevant to syntax as a whole. Would including other parameters bring about differences in model dynamics? Along these two important directions, progress will be possible once a significantly extended PCM database is made available.

Progress that depends on computational modeling work, instead, involves replacing the random assignment of hidden weights with an educated guess based on the available statistics of parameter values. This is called `network reconstruction', and much effort has been invested in developing it, for example in the context of protein folding \cite{morcos2011direct} or recording the activity of multiple single neurons \cite{roudi2009ising}. Appropriate methods are still to be explored in the case of syntax, which would seem very straightforward, were it not for the implications. With the random assignment, natural languages turn out to be far from any of the multiple quasi-stable configurations of parameter values the network rapidly approaches in the first few time steps, as those configurations depend on hidden weights that bear no relation to existing languages. This prevents making any form of detailed predictions about the evolution of the syntax of individual languages. The hope, then, is that effective strategies for network reconstruction will eventually lead to specific predictions or accounts of specific instances of diachronic change.

An even more ambitious goal is to make inferences about parameter couplings from electrophysiological (EEG/MEG) data, neuropsychological observations in patients with language impairments or studies of language acquisition in children. While the first two seem remote possibilities at the moment, it is encouraging to see the beginning of a structured understanding of the sequential processes of parameter acquisition \cite{friedmann2021growing} (but see \cite{bosch2025another}). The emerging sequential patterns, once generalized across a significant number of languages, represent a form of partial ordering that can be combined with the ordering associated with implications, as in Fig.\ref{implications}; further, parameters should be distinguished depending on the type of their putative default state \cite{longobardi2023grammatical,crisma2025your} and on the ease with which each parameter can be set, by a speaker acquiring the language, from a corpus of spoken utterances \cite{crisma2025your}. These different lines of evidence can aid the reconstruction of a less random assignment of hidden weights, to ultimately bring the model closer to real language phenomenology, as we plan to pursue in future work.

\vspace{4pt}

\textbf{Acknowledgments.} 
We are indebted to the many researchers who have contributed to the PCM database, foremost Cristina Guardiano and Paola Crisma; and to Eric De Giuli for stimulating discussions.  
	
\vspace{4pt}

\textbf{Data Availability.} 
The PCM Database has been published, e.g. in \cite{ceolin2021boundaries} and \cite{crisma2025your} and expanded versions are likely to be published in the near future. The software used in the present study will be made freely available upon publication.

\vspace{4pt}

\textbf{Author Contributions.} 
H.Y. carried out the numerical experiments and drafted the manuscript; K.I.R. contributed to the model formulation and carried out early simulations; G.L. conceptualized the study, structured the syntactic implications in the PCM database and revised the manuscript; A.T. conceptualized the study, coded the implications as polysynaptic weights and drafted the manuscript.

	\bibliography{refs} 

\appendix
\renewcommand\thefigure{\thesection.\arabic{figure}}    
\section{Syntactic dependence}
\setcounter{figure}{0}  

\begin{figure*}
	\centering
	\includegraphics[width=13cm]{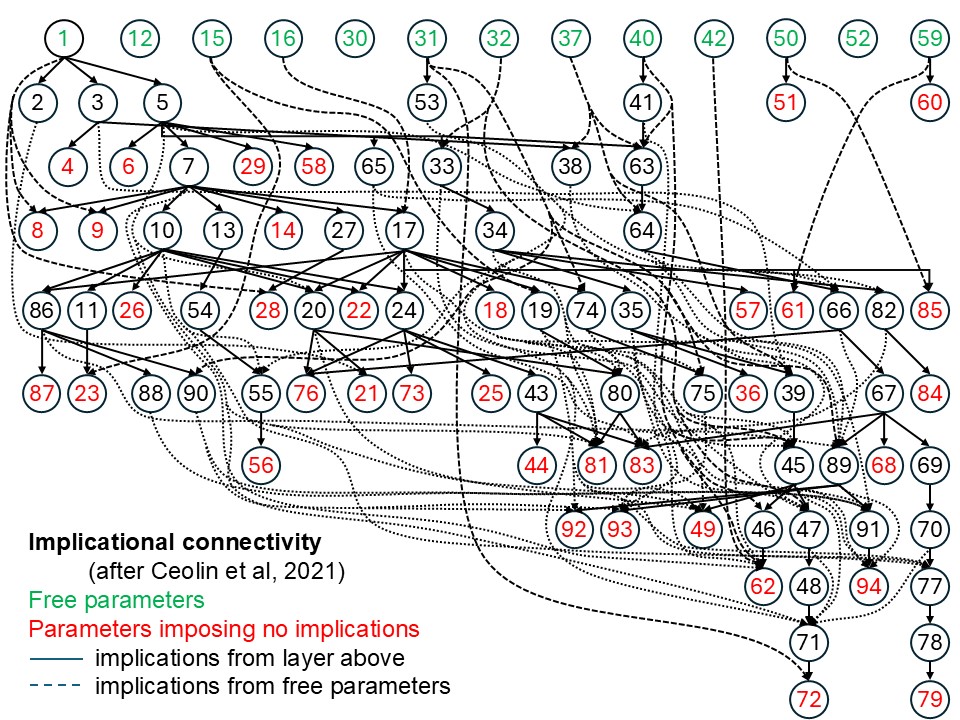}
	\caption{\textbf{Syntactic implications operate at different depth levels.}
The implications are shown as arrows pointing at the dependent parameters, numbered as in the PCM database (see \cite{ceolin2021boundaries}; level $L_k$ with $k=0,\dots , 10$ are arranged top to bottom, while the number $k'$ of arrows impinging on each parameter correspond to its class $D_{k'}$, with $k'=0,\dots ,6$. Boolean rules are not represented in this simple connectivity diagram, but are detailed in their Suppl. Table 2). Note that $L_0\equiv D_0$.} 
	\label{implications}
\end{figure*}

To clarify how implications constrain changes in the values of the syntactic parameters, we could proceed down the levels of Fig.\ref{implications}, but it is somewhat more perspicuous to group them into classes according to their dependence degree — that is, the number of other parameters their relevance depends upon — labeling these groups \( D_0, D_1, D_2, \ldots, D_6 \). For example, a parameter in $D_3$ is only relevant if 3 other parameters take specific values; otherwise it is undefined, which in the first version of our model is conflated with taking the default value -1, in the second it is assigned value 0.

\begin{figure*}
	\centering
	\includegraphics[width=15cm]{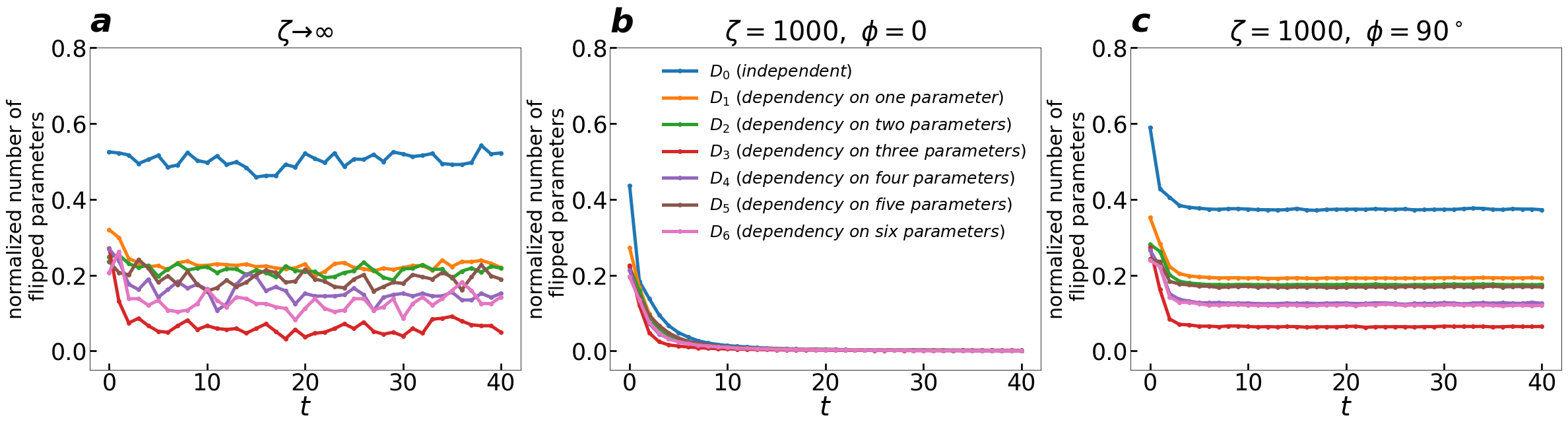}
	\caption{\textbf{Dynamical evolution of syntactic parameters grouped by dependence number.}
\textbf{(a)} Temporal evolution of the normalized number of flipped parameters for each class \( D_0, D_1, \ldots, D_6 \) in the deterministic limit (\( \zeta \to \infty \)), in the first version of the model. The 13 independent parameters in class \( D_0 \), which are unconstrained by implications, are the most active, flipping around half the times. 
\textbf{(b)} Same as (a), but with added weak symmetric hidden interactions (\( \zeta = 1000 \), \( \phi = 0^\circ \)). They suffice to freeze the system of parameters.
\textbf{(c)} Same as (a), but with weak {\em antisymmetric} stochastic interactions (\( \zeta = 1000 \), \( \phi = 90^\circ \)), which keep the system fluid.} 
	\label{figS1}
\end{figure*}

First, we look at network dynamics in the deterministic limit (\( \zeta \to \infty \)), where dynamics are governed purely by implications, in the absence of the extra interactions, which in our model are randomly assigned. The model is the purely binary one. Figure~\ref{figS1}\textbf{a} shows the temporal evolution of the normalized count of flipping parameters for each syntactic class. Independent parameters, the core set of 13 syntactic features in  \( D_0 \) not dependent on implications, of course flip every second time step, on average. Changes in these free parameters propagate through the hierarchy, making parameters  in $D_1$ those with the next highest probability to flip, then those in $D_2$, and so on. Ultimately, therefore, all syntactic changes are enabled, in this limit, by flipping the 13 independent parameters, out of the total 94, in the absence of hidden couplings. When the latter are introduced, even with minimal strength (\( \zeta = 1000 \), see Fig.~\ref{figS1}\textbf{a} and \textbf{b} for \( \phi = 0^\circ \) and \( \phi = 90^\circ \), respectively), the free parameters are not free anymore: for them, unconstrained by implications, the hidden couplings are the interactions there are, and they suffice to freeze the evolution of the free parameters in a few time steps, if symmetric or nearly symmetric. The other parameters then freeze as a result, see Fig.~\ref{figS1}\textbf{a} for \( \phi = 0^\circ \). Beyond approximately \( \phi = 30^\circ \) the hidden interactions are asymmetric enough to keep the system fluid, and the fluidity is expressed in a similar hierarchy of flipping rates as the one seen in the absence of hidden couplings (as Fig.~\ref{figS1}\textbf{c} shows for \( \phi = 90^\circ \)).

\end{document}